\documentclass[Royal,sageh,times]{sagej}

\usepackage{moreverb,graphicx}
\usepackage[hyphens]{url}
\usepackage{float}

\usepackage[colorlinks,bookmarksopen,bookmarksnumbered,citecolor=red,urlcolor=red]{hyperref}

\usepackage{caption}
\usepackage{subcaption}

\newcommand\BibTeX{{\rmfamily B\kern-.05em \textsc{i\kern-.025em b}\kern-.08em
T\kern-.1667em\lower.7ex\hbox{E}\kern-.125emX}}
\def\volumeyear{2022}

\begin{document}


\runninghead{Vierø, Vybornova \& Szell}

\title{BikeDNA: A Tool for Bicycle Infrastructure Data \& Network Assessment}

\author{Ane Rahbek Vierø\affilnum{1}, Anastassia Vybornova\affilnum{1}, Michael Szell\affilnum{1,}\affilnum{2,}\affilnum{3}}

\affiliation{\affilnum{1}IT University of Copenhagen, Denmark\\
\affilnum{2}ISI Foundation\\
\affilnum{3}Complexity Science Hub Vienna\\}

\corrauth{Ane Rahbek Vierø, Computer Science Department,
IT University of Copenhagen,
}

\email{anev@itu.dk}

\begin{abstract}
High-quality data on existing bicycle infrastructure are a requirement for evidence-based bicycle network planning, which supports a green transition of human mobility. However, this requirement is rarely met: Data from governmental agencies or crowdsourced projects like OpenStreetMap often suffer from unknown, heterogeneous, or low quality. Currently available tools for road network data quality assessment often fail to account for network topology, spatial heterogeneity, and bicycle-specific data characteristics. To fill these gaps, we introduce BikeDNA, an open-source tool for reproducible quality assessment tailored to bicycle infrastructure data with a focus on network structure and connectivity. BikeDNA performs either a standalone analysis of one data set or a comparative analysis between OpenStreetMap and a reference data set, including feature matching. Data quality metrics are considered both globally for the entire study area and locally on grid cell level, thus exposing spatial variation in data quality. Interactive maps and HTML/PDF reports are generated to facilitate the visual exploration and communication of results. BikeDNA supports quality assessments of bicycle infrastructure data for a wide range of applications -- from urban planning to OpenStreetMap data improvement or network research for sustainable mobility.
\end{abstract}


\maketitle

\section{Introduction}

Cities across the globe are striving to make their transportation systems more environmentally and socially sustainable. One of the most cost-effective ways to do so is to boost cycling, like in Paris or Bogotá \citep{c40_cities_upgrade_2019, city_of_paris_nouveau_2021}, and as recommended by the European Commission and the IPCC \citep{ec_new_2021, jaramillo_transport_2022}. However, achieving a substantial shift towards cycling remains a challenge. Adequate infrastructure is typically lacking, as is the \emph{collection and provision of data} on bicycle infrastructure, which is necessary to harness the potential of data-driven bicycle network planning. Recent advances demonstrate that quantitative analyses of bicycle infrastructure on the network level can assist planning decisions and considerably improve the impact of planned investments \citep{natera_orozco_data-driven_2020, olmos_data_2020, steinacker_demand-driven_2022, szell_growing_2022, vybornova_automated_2021}, but this necessitates complete and high-quality infrastructure data. Moreover, even when such data are available, there is often little knowledge on data completeness and quality. This is true both for administrative data and for crowdsourced data such as OpenStreetMap (OSM) \citep{nelson_crowdsourced_2021, ramboll_walking_2022}.

\begin{figure}[t]
\centering
\includegraphics[width=0.55\textwidth]{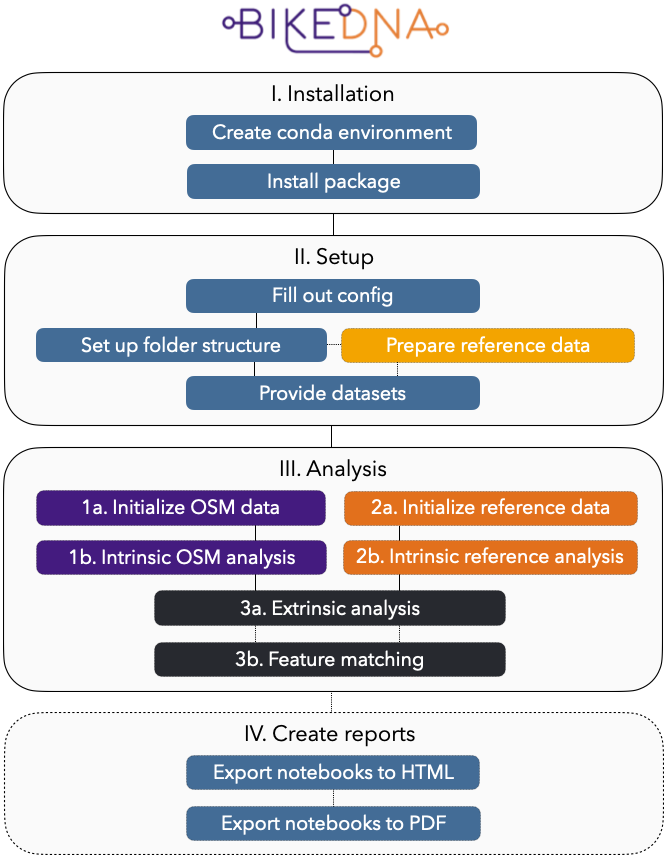}
\caption{The pipeline of BikeDNA. The analysis (III) is divided into three parts: 1) Intrinsic analysis of OSM bicycle network data (purple), 2) intrinsic analysis of reference bicycle network data (orange), and 3) and comparison of OSM and reference data (black), including feature matching. Dotted parts are optional.}
\label{fig:pipeline}
\end{figure}

To fill this gap and to help researchers, planners, and others in the field assess the quality of bicycle infrastructure data, we introduce \href{https://github.com/anerv/BikeDNA}{BikeDNA} (Bicycle Infrastructure Data \& Network Assessment). BikeDNA is a computational tool written in Python that performs a reproducible quality assessment on one or two bicycle infrastructure data sets (see Fig.~\ref{fig:pipeline}). `Bicycle infrastructure' is defined in BikeDNA as any part of the road network that is designated specifically for cycling; this definition can be modified by the user. BikeDNA can perform standalone analysis of either OSM or administrative data. If both OSM and administrative data are available, BikeDNA compares differences between the two data sets. To our knowledge, there are currently no other open-source tools for data quality assessment available that are tailored to the specific nature of bicycle infrastructure data.

\subsection{Previous research on spatial data quality}

While BikeDNA is designed to be compatible with both crowdsourced and administrative data on bicycle infrastructure, the quality assessment of OSM data is a central feature of the tool due to the widespread usage of OSM for bicycle network research \citep{ferster_using_2020, nelson_crowdsourced_2021}. Much recent work on spatial data quality has moreover focused on the quality of crowdsourced or volunteered data, in attempts to build trust in data sets from non-official sources. Our review of previous work is therefore focused on quality assessment of crowdsourced road network data. For more comprehensive, in-depth reviews of quality assessment methods for Volunteered Geographic Information (VGI), we refer to  \cite{fonte_assessing_2017}, \cite{senaratne_review_2017}, and \cite{degrossi_taxonomy_2018}. For an overview of solutions for spatial data quality more generally, see \cite{medeiros_solutions_2019}.

\subsubsection{Intrinsic versus extrinsic analysis}

\begin{figure}[t]
\centering

\includegraphics[width=0.999\textwidth]{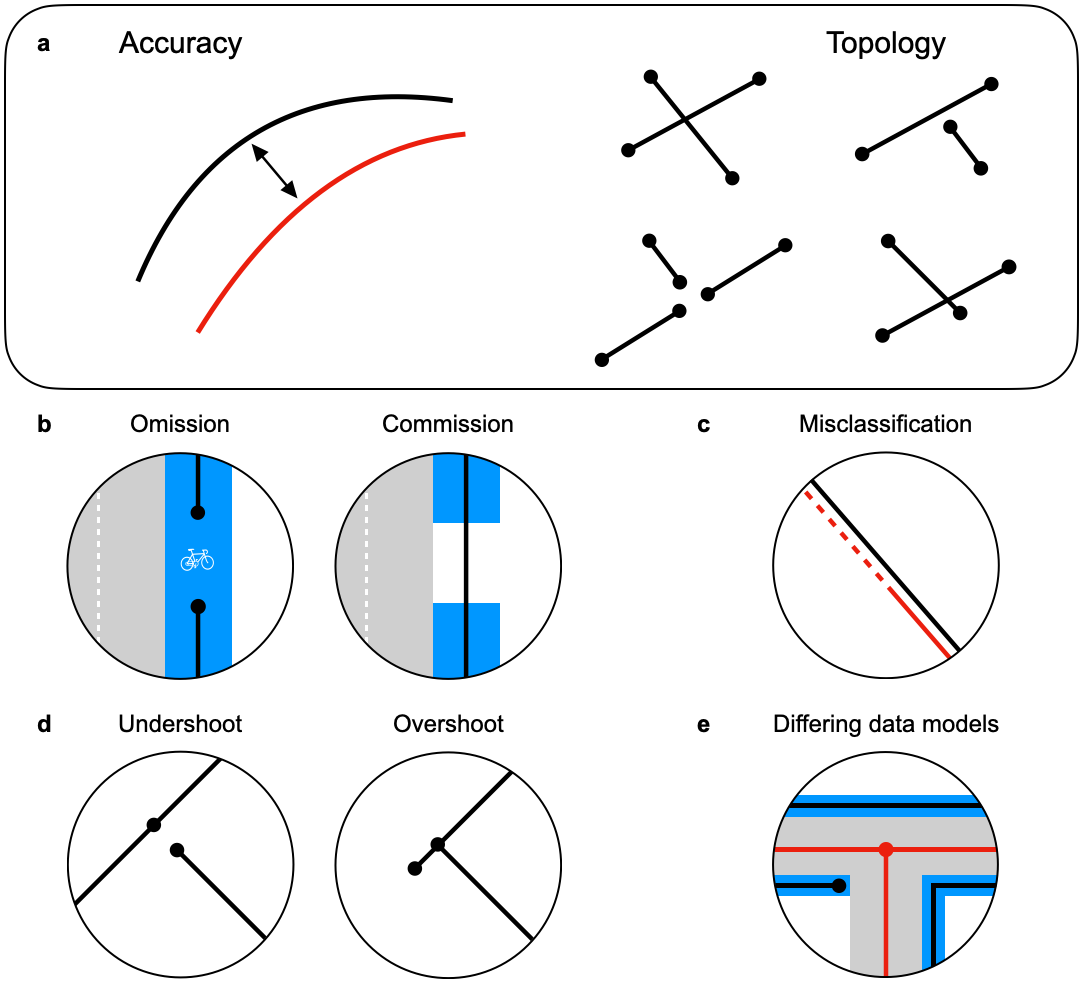}
\caption{Known quality issues in bicycle infrastructure data. a) Different aspects of data quality assessment: Accuracy (left) versus topology (right). Early research on spatial data quality focused on accuracy, while BikeDNA puts a special emphasis on topology. Adapted from \cite{haklay_how_2010} and \cite{neis_street_2012}. b) Errors of omission (left) or commission (right) result in differing data completeness. c) Misclassification leads to different types of bicycle infrastructure in two corresponding data sets (red versus black). d) Edge undershoot (left) and edge overshoot (right). e) Mapping of bicycle infrastructure to the centerline of the road (red) or on the side (black). In this case, the centerline mapping creates routable network data at the cost of accuracy.}
\label{fig:knownissues}
\end{figure}

A common way to classify spatial data quality assessment methods is as either `intrinsic' (standalone) or `extrinsic' (comparative) -- a classification also used by BikeDNA. Intrinsic methods study inherent properties of the data set itself or how it was created, while extrinsic methods make use of an external reference data set for comparison. Early work on VGI data quality was primarily occupied with extrinsic evaluations with an emphasis on evaluating the completeness and spatial accuracy, using local metrics for the length of the mapped network in an area \citep{haklay_how_2010-1} as well as more advanced methods for feature matching \citep{koukoletsos_assessing_2012}. While data completeness and accuracy are relevant data quality metrics, they do not address topological quality, i.e.~how network elements are connected \emph{structurally}, which however is crucial for many applications of road network data (see Fig.~\ref{fig:knownissues}). For this reason, extrinsic approaches were quickly expanded to also include road network topology, routing, and other network attributes \citep{girres_quality_2010, mondzech_quality_2011, neis_street_2012, zielstra_using_2012, graser_towards_2014}. Intrinsic evaluations of VGI data quality have been developed for situations where no reference data are available, or to avoid some of the more computationally intensive methods used in extrinsic comparisons. Methods for intrinsic evaluation analyze the spatial and structural dimensions of the data  \citep{neis_street_2012, barron_comprehensive_2014}, as well as the number of contributors and edits \citep{kesler_tracking_2011, neis_comparison_2013, barron_comprehensive_2014, grochenig_estimating_2014}. Extrinsic and intrinsic methods are often used in combination, sometimes in more elaborate frameworks for assessing spatial data quality \citep{barron_comprehensive_2014, fabrikant_conceptual_2015}, as it is the case for BikeDNA.

\subsubsection{Quality standards \& fitness for purpose}

Another way to classify existing approaches to spatial data quality is to distinguish between the formulation of standards for spatial data quality versus the idea of `fitness for purpose'. Following the former approach, for example, \href{https://www.iso.org/standard/32575.html}{ISO 19157} defines spatial data quality as composed of: accuracy (positional, thematic, and temporal); data completeness; internal logical consistency; and finally, usability \citep{fonte_assessing_2017}. While these are useful concepts that are used throughout the quality assessment in BikeDNA, the consistency implied by formal standards for data quality is difficult to apply to heterogeneous, volunteered data like OSM \citep{hashemi_assessment_2015}, and might be less relevant for some OSM use cases. Therefore, \cite{barron_comprehensive_2014} suggest to instead evaluate OSM data based on fitness for purpose, meaning that the quality and suitability of a data set are evaluated based on whether the data fulfill the requirements for the individual use case. This is also the approach adopted in BikeDNA. Therefore, instead of referencing universal standards, BikeDNA is designed to help decide whether a data set is \textit{good enough} for an intended use case.

\subsubsection{Gaps in network-level bicycle data assessment and tools}

Only few studies of quality assessments specifically of bicycle infrastructure data exist, with \cite{hochmair_assessing_2015} and \cite{ferster_using_2020} as the most prominent examples. Both studies compare OSM data on dedicated bicycle infrastructure to external reference data sets, assessing data completeness generally and for different types of bicycle infrastructure. This is of particular interest, since studies on the general completeness of OSM have shown that bicycle lanes and paths often are among the later features to be mapped \citep{neis_street_2012, barron_comprehensive_2014}. These studies provide important insights into the completeness of bicycle infrastructure data in OSM, but they do not address more complex measures of spatial data quality, such as network topology. This is a highly relevant research gap, since infrastructure data for pedestrians and cyclists are more prone to topological errors than data for motorized transport \citep{neis_street_2012}. Many of the assumptions used to evaluate road networks for car traffic, e.g.~interpreting disconnected components and small gaps between edges as errors, do furthermore not hold for bicycle networks which are often built in a fragmented and piece-wise manner \citep{natera_orozco_data-driven_2020, szell_growing_2022}.

Overall, the quality of VGI and OSM data is well-studied, particularly when it comes to data on the car road network \citep{fonte_assessing_2017}, and OSM data are often shown to be of high quality compared to other data sources \citep{neis_street_2012, zielstra_using_2012, hochmair_assessing_2015, ferster_using_2020, zhang_detecting_2021}. However, data errors and inconsistencies are not randomly distributed in OSM, but instead correlate with lower population densities, socioeconomic variables, and vary from country to country \citep{haklay_how_2010-1, ferster_using_2020}. Administrative data have also been shown to suffer from uneven data quality, for example due to the role of local data maintainers and differing mapping practices \citep{hvingel_gode_2023}. For this reason, quality assessment in one particular location cannot be generalized to other locations. Moreover, the quality of data on infrastructure for cyclists often lags behind the quality of other road network data, both in terms of completeness and consistency. Although these findings have profound implications for bicycle research that uses OSM data, to date only few studies have assessed specifically the quality of bicycle infrastructure data in OSM. In short, the lack of studies and tools that assess how well bicycle infrastructure has been mapped is a considerable barrier to OSM data uptake and to data-driven bicycle network planning.

\subsection{Filling the gap with BikeDNA} 

To addresses this gap of know-how and lack of available methods for quality assessment, we have developed BikeDNA. The tool contains a systematic and easy-to-follow quality assessment of bicycle infrastructure data. This is relevant for use cases such as bicycle routing \citep{murphy_implementing_2019}, connectivity analysis \citep{natera_orozco_data-driven_2020, vybornova_automated_2021}, network quality assessment \citep{dill_measuring_2004}, and accessibility studies \citep{peopleforbikes_bna_2023}. BikeDNA implements and visualizes a wide range of existing spatial VGI and OSM data quality metrics, but is tailored specifically to bicycle infrastructure data. The main features of BikeDNA are:
\begin{itemize}
  \item Analysis of completeness, network density, and OSM tags
  \item Analysis of network topology and connectivity
  \item Feature matching between OSM and reference data 
  \item Creation of detailed PDF and interactive HTML reports
\end{itemize}

\noindent BikeDNA thus addresses both data \textit{completeness}, i.e.~how much information each data set contains, and data \textit{consistency} and \textit{accuracy}. BikeDNA moreover accounts for different scenarios of data availability and can be applied in scenarios where only OSM, only reference data, or both are available. All data quality metrics, whenever relevant, are computed globally for the entire study area and locally for each cell in a grid covering the study area, which allows to detect and visualize local variations in data quality. The quality metrics are made accessible by extensive documentation and interpretation assistance in each step, and automated plotting of results with static and interactive maps. BikeDNA can help improve the quality of a data set by pinpointing both in which \emph{features} and which \emph{locations} gaps, errors, or inconsistencies exist. Potential users are planners, researchers, data maintainers, and everyone else who needs an indication of data quality relevant to bicycle applications.

The rest of the paper is organized as follows: In the next section, we briefly discuss the conceptual and analytical framework of BikeDNA. Then, we provide a descriptive overview of the tool: First, we show how to use BikeDNA, then we provide example outputs of BikeDNA's quality assessment of OSM and the open government data `GeoDanmark' for a showcase area in Greater Copenhagen, Denmark. We conclude with a discussion of limitations and the need for future work on bicycle data quality. In addition, we outline possible improvements to BikeDNA and quality assessments more generally. The full description and technical specifications of BikeDNA are available on GitHub: \href{https://github.com/anerv/BikeDNA}{https://github.com/anerv/BikeDNA}.

\section{Evaluating bicycle infrastructure data}

In the following section, we introduce the conceptual and analytical framework with an overview of the different scenarios of data availability and data sources considered by BikeDNA, describe known data quality issues in bicycle infrastructure data, and discuss how to perform and interpret quality assessments without ground truth data.

\subsection{Data sources}

BikeDNA considers three different scenarios of data availability for bicycle networks: either data are available from OSM; data are available from an administrative source, e.g.~a local municipality; or data are available from both. OSM is the primary data source for data used in bicycle research and routing applications. To our knowledge, administrative data are used primarily by governmental institutions. OSM and administrative data are not necessarily at odds, since OSM data sometimes are partly based on administrative data sources \citep{zielstra_assessing_2013, ramboll_walking_2022}. 

Bicycle infrastructure data from OSM are in many cases of a similar or even higher quality than administrative data when it comes to aspects such as data completeness \citep{hochmair_assessing_2015, ferster_using_2020}. This is partly due to a lack of government resources, as well as the unfavorable treatment of active mobility in  data collection and maintenance in many jurisdictions \citep{ramboll_walking_2022}. The inherently heterogeneous nature of VGI data, combined with the lack of metadata on data quality parameters, does however pose a barrier to data uptake in, for example, public transport planning \citep{hashemi_assessment_2015}. The lack of standardized quality assurance for crowdsourced data necessitates methods for local data quality assessment, which however rarely are available. BikeDNA seeks to remediate this gap.

\subsection{Known quality issues}

Within the fitness for purpose approach, it depends on the use case whether certain data errors and inconsistencies are considered a problem. For example, spatial inaccuracies such as smaller displacements of objects are rarely an issue for network-based bicycle research, as long as the data topology is correctly represented (see Fig.~\ref{fig:knownissues}a). There are, however, several types of errors and inconsistencies which will be a problem for most data applications, and which we briefly explain below.

\subsubsection{Errors of omission/commission} `Errors of omission' describe missing data, while `errors of commission' refer to features in the data set that should not be there (see Fig.~\ref{fig:knownissues}b). These error types appear in both OSM and administrative data sets on bicycle infrastructure \citep{hochmair_assessing_2015, ferster_using_2020}.

\subsubsection{Misclassification} Misclassification happens when bicycle infrastructure is classified as something else, or when non-bicycle infrastructure is classified as bicycle infrastructure, resulting in errors of omission and commission, respectively (see Fig.~\ref{fig:knownissues}c). For example, unprotected bicycle infrastructure might be misclassified as protected, which has implications for example for Levels of Traffic Stress \citep{mekuria_low-stress_2012}. Different types of misclassification errors have been documented in previous research on the quality of OSM bicycle infrastructure data \citep{hochmair_assessing_2015, ferster_using_2020}.

\subsubsection{Topology errors} Topology errors occur where network edges are not properly connected at intersections, either because of a missing node at intersections, or in situations of ‘undershoots’ (one or more edges are too short and thus do not connect at an intersection) or ‘overshoots’ (one or more edges are too long at an intersection, which creates small edges with dangling nodes), as shown in Fig.~\ref{fig:knownissues}d. Within OSM, this type of problem is more prominent for bicycle infrastructure than for other types of the road network \citep{neis_street_2012}.

\subsubsection{Inconsistent mapping procedures} \label{subsec:mapping_methods} While not technically an error, differing or inconsistent mapping methods make it difficult to assess data completeness. Within some approaches, bicycle lanes are digitized as road center lines, regardless of whether only one or both sides of the road have a bicycle lane. Other approaches digitize bicycle lanes as their individual geometries, in which case bicycle lanes on two sides of the same road are represented separately in the data set (see Fig.~\ref{fig:knownissues}e for an illustration). This can significantly distort measurements of network density and network length and makes it complicated to compare data sets that are based on differing mapping approaches. Other studies have approached this challenge by only including one lane in length computations \citep{hochmair_assessing_2015} or by generalizing parallel lines to the center line \citep{ferster_using_2020}. 

\subsection{Quality assessment without ground truth data}

The absence of ground truth or authoritative data sets on bicycle infrastructure is a challenge -- not only for bicycle planners and researchers, but for anyone attempting to assess data quality. Without validation data, data quality assessment cannot be fully automated. Moreover, given the fragmented, low-quality nature of many actual bicycle networks, it is often impossible to distinguish in an automated way whether a poor performance of a data quality metric is due to low \emph{data} quality or due to low quality of the \emph{infrastructure} itself. For these reasons, BikeDNA does not issue any final verdict about the quality of the analyzed data sets. Rather, it allows for exploring different characteristics of the data which, in combination with local knowledge, can help make a better-informed assessment of the data quality and potential limitations to the data usability. As pointed out by \cite{brovelli_towards_2017} and \cite{ferster_using_2020}, local knowledge of both on-the-ground conditions and OSM mapping norms is crucial for verification and interpretation of quality assessments. Accordingly, BikeDNA emphasizes the relevance of local knowledge and familiarity with the area for any kind of planning or research process.

\section{How BikeDNA works}

Here, we briefly explain how to use BikeDNA. The pipeline follows four main steps: I.~Installation, II.~Setup, III.~Analysis, and IV.~Creation of reports (optional), see Fig.~\ref{fig:pipeline}. For all details on specifications, installation and setup, we refer to the \texttt{README} and the explanations in the notebooks, all found on the \href{https://github.com/anerv/BikeDNA}{GitHub repository}. BikeDNA runs entirely in Python via Jupyter notebooks. Therefore, knowledge of handling Python Jupyter notebooks, but no programming proficiency, is required for using BikeDNA. BikeDNA makes extensive use of open-source Python libraries, most importantly \texttt{GeoPandas} \citep{jordahl_geopandasgeopandas_2021}, \texttt{OSMnx} \citep{boeing_osmnx_2017}, \texttt{NetworkX} \citep{hagberg_exploring_2008}, and \texttt{Momepy} \citep{fleischmann_momepy_2019}. 

Once the Python environment is created (step~I), the configuration file is filled out and a potential reference data set is prepared (step~II), the entire analysis is conducted by running the Jupyter notebooks (step~III). The notebooks are structured as follows: preparation and intrinsic analysis of OSM  data (\textit{1a, 1b}); preparation and intrinsic analysis of reference data (\textit{2a, 2b}); extrinsic analysis of OSM and reference data (\textit{3a}); and finally, feature matching between OSM and reference data (\textit{3b}). The intrinsic analyses of OSM and reference data can be run independently, while the extrinsic analysis requires prior intrinsic analyses of both data sets.

The optional last step IV is the generation of detailed HTML (including interactive maps) or PDF reports of BikeDNA's  results, based on the executed notebooks. All outputs of BikeDNA (processed data, graph objects, figures, static and interactive maps, data identified as inconsistent or erroneous) are automatically saved in corresponding subfolders during notebook execution.

BikeDNA has a few required user inputs:
\begin{enumerate}
    \item \textbf{Study area.} A polygon defining the study area to be analyzed.
    \item \textbf{OSM queries.} BikeDNA comes with default queries for defining the bicycle network based on the OSM data, which are automatically downloaded as part of the OSM analysis. If OSM analysis is performed, the queries should be adapted to account for local differences in tagging practices and local definitions of what types of infrastructure the bicycle network includes.
    \item \textbf{Reference data.} If a reference data set is provided, the data must be pre-processed according to the \href{https://github.com/anerv/BikeDNA/blob/main/datasetrequirements.md}{data set requirements} (SI file 2). Additionally, settings for how to parse the reference data attributes must be provided in the configuration file.
\end{enumerate}
Apart from these required inputs, BikeDNA also allows for an optional user configuration of its analysis parameters, such as distance thresholds used in identification of over/undershoots, OSM tags to include in the missing tag analysis, and distance thresholds used for feature matching. 

\section{What BikeDNA does}

In this section, we present the main features and outputs of BikeDNA by the example of an area in Greater Copenhagen, using a local data set from GeoDanmark \citep{geodanmark_2023} as the reference data. Detailed elaborations of methodology and interpretation of results are found in the analysis notebooks (see SI file 1). All code and results can also be found on the `GeoDanmark' branch on the \href{https://github.com/anerv/BikeDNA/tree/GeoDanmark}{GitHub} repository.

\begin{figure}[t]
\centering
\includegraphics[width=0.999\textwidth]{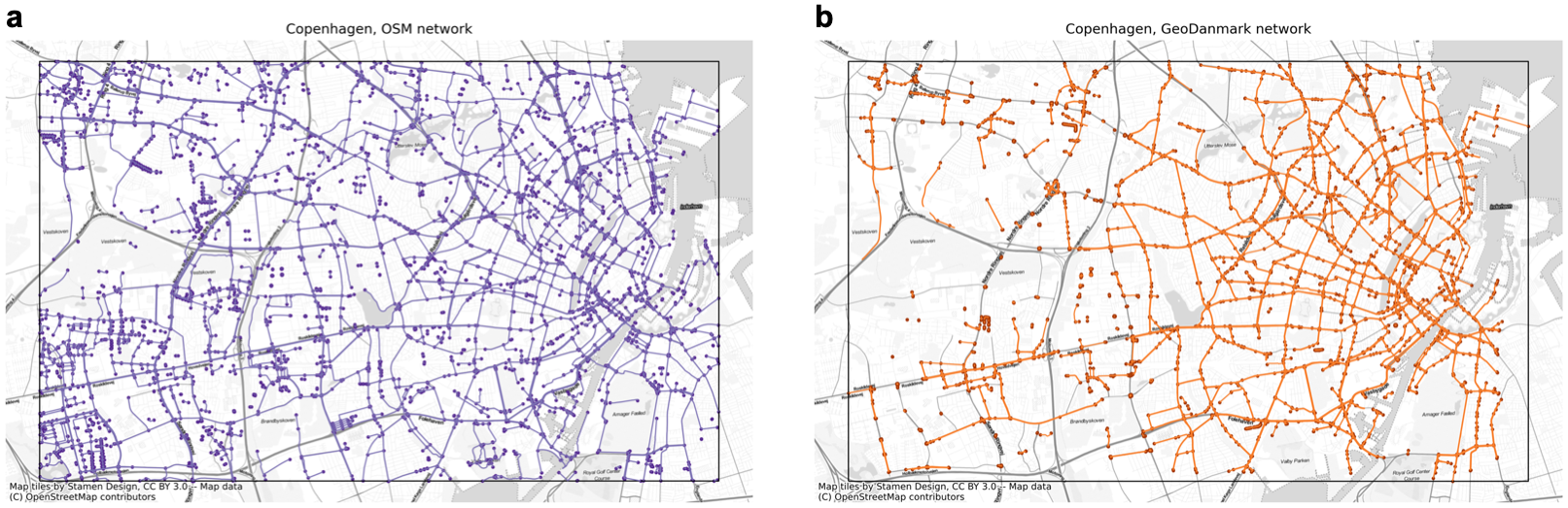}
\caption{Input data for test area from a) OSM and b) GeoDanmark. Maps created with BikeDNA v.1.0.0.}
\label{fig:inputdata}
\end{figure}

\subsection{Global \& local analysis} 
BikeDNA conducts analysis on two levels: global and local. In global analysis steps, aggregated metrics are computed for the entire study area. Some results, such as the total network length, are only meaningful on a global scale. For some data inconsistencies, however, we do not only want to know \textit{whether} they occur, but also \textit{where} they happen. Their location is particularly relevant because of the spatial heterogeneity and non-random patterns in locations of high and low data quality that often characterize crowdsourced data \citep{haklay_how_2010-1}. Therefore, in the local analysis steps, BikeDNA computes metrics separately for each grid cell on a customizable square grid that covers the study area.

\subsection{Network density} The density of a transportation network is defined as the length of edges or number of nodes per square kilometer, which is the most basic descriptive statistic that can indicate data completeness. Comparing completeness between two bicycle infrastructure data sets based on network density is notoriously difficult due to differing mapping approaches and data models (see Fig.~\ref{fig:knownissues}e). OSM implements both methods for mapping bicycle infrastructure, as discussed in the section \nameref{subsec:mapping_methods} above. Therefore, BikeDNA computes the edge density based on the infrastructure length, \emph{not} geometric edge length. For example, a 100-meter-long bidirectional path is counted as 200 meters of bicycle infrastructure. This allows to compare data completeness between data sets with differing mapping approaches. 

\begin{figure}[t]
\centering
\includegraphics[width=0.999\textwidth]{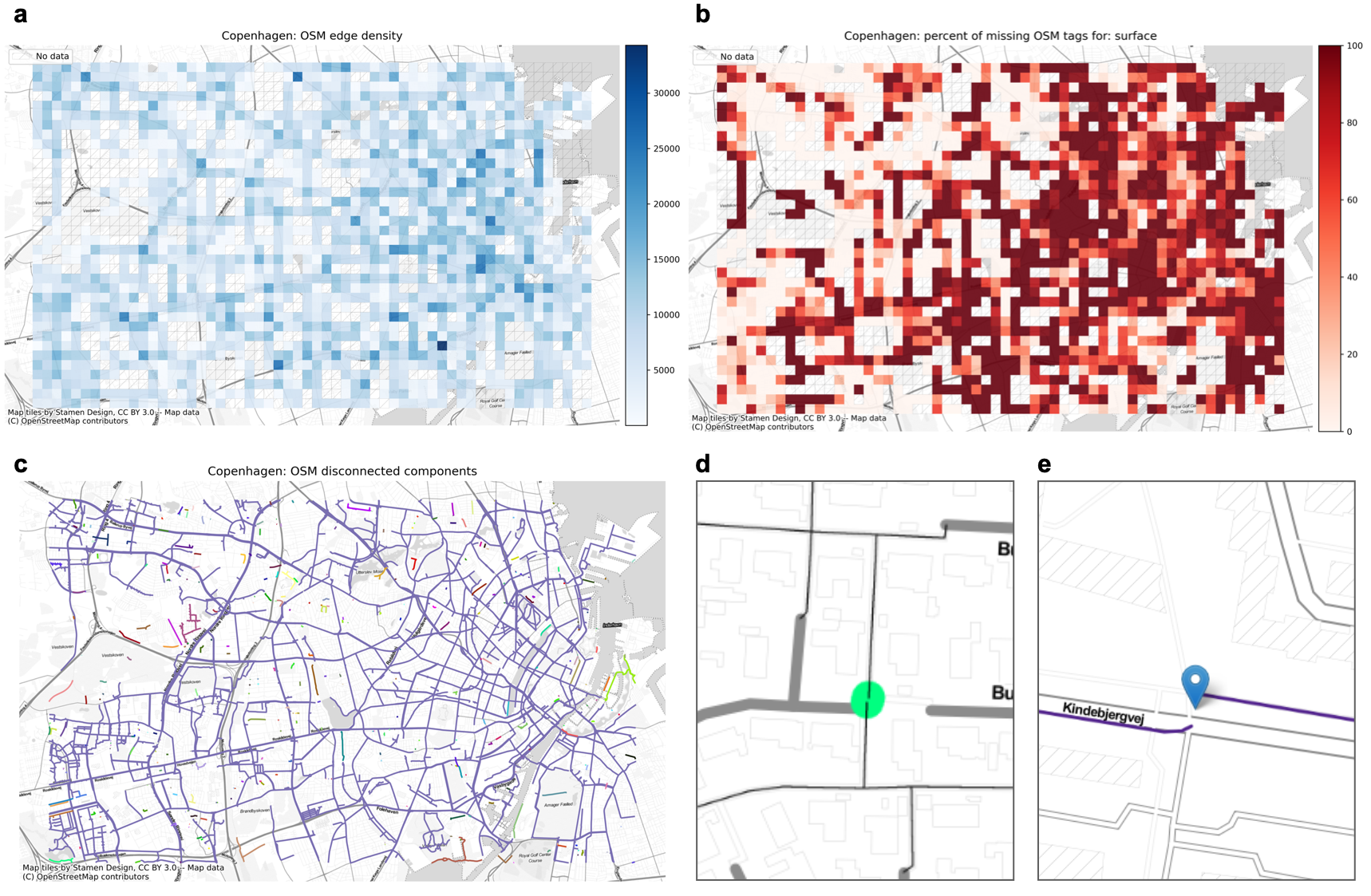}
\caption{Results from intrinsic analysis of OSM and GeoDanmark data: a) OSM edge density. b) Missing OSM tags. c) Disconnected components. d) Example of an undershoot. e) Edges from two disconnected components with less than the specified distance threshold between them. Maps created with BikeDNA v.1.0.0.}
\label{fig:intrinsic}
\end{figure}

The analysis requires a simplified network in order not to count interstitial network nodes (i.e.,~nodes that do not represent intersections or dead-ends) in the computation of network density. For this reason, both OSM and reference networks are simplified using a modified OSMnx function \citep{sebastiao_modified_2022} for network simplification, which keeps nodes only at intersections and dead-ends, as well as at locations where the value of important attributes changes.

Within the intrinsic analysis, network density values can indicate under- or overmapped regions of the study area, if the density pattern strongly deviates from expected patterns. In the extrinsic analysis, contrasting network density values for OSM and the reference data set can be used for comparing data completeness \citep{haklay_how_2010-1}. 

\subsection{Consistency of OSM tags}

One important characteristic of OSM data on bicycle infrastructure is the highly heterogeneous distribution of tags (e.g.~width, speed limits, street lights, or other characteristics of interest to cyclists), which poses a barrier to evaluations of e.g. bikeability \citep{wasserman_evaluating_2019}. Likewise, the lack of restrictions on OSM tagging can lead to conflicting tags, which undermines the evaluation of bicycle conditions. For this reason, BikeDNA allows for checking the consistency of tags in OSM, tailored to OSM’s data structure. BikeDNA conducts analysis of OSM tags in three ways: by identifying and visualizing where user-defined OSM tags are lacking information; by highlighting where edges are labelled with two or more tags defined as contradictory by the user; and finally, by visualizing tagging patterns, i.e.~the spatial variation in tags that are used to describe bicycle infrastructure in OSM. In addition, BikeDNA makes use of OSM tags to identify missing intersection nodes: when two edges intersect without having a node at the intersection and neither of them is tagged as bridge or tunnel, it is a clear indication of a topology error \citep{neis_street_2012, barron_comprehensive_2014}. 

\begin{figure}[t]
\centering
\includegraphics[width=0.999\textwidth]{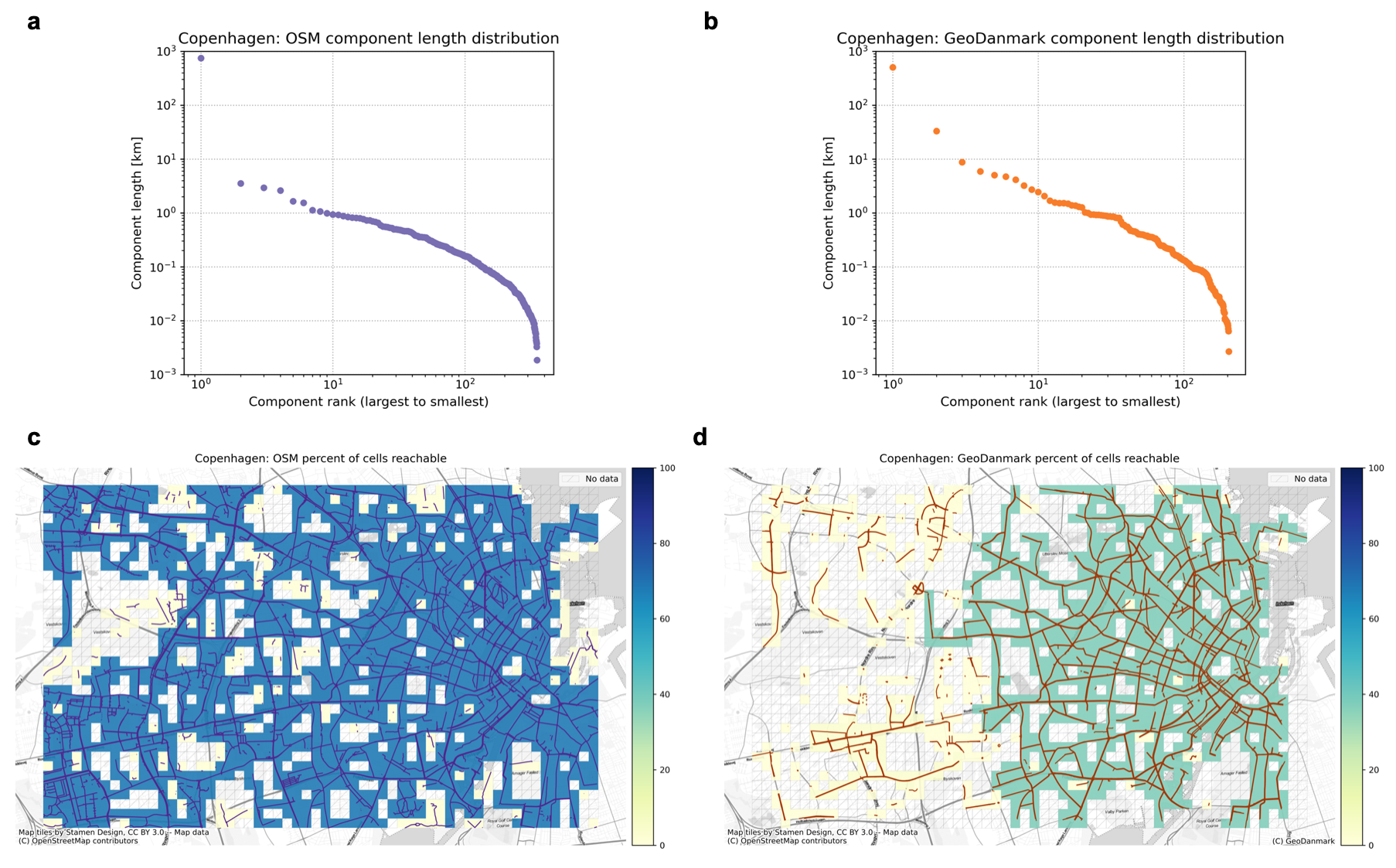}
\caption{Extrinsic analysis based on a comparison of intrinsic results: Zipf plots of a) OSM and b) GeoDanmark component length distribution. Percent of cells reachable through c) the OSM and d) the reference networks. Maps created with BikeDNA v.1.0.0.}
\label{fig:extrinsic}
\end{figure}

\subsection{Network topology: Under/overshoots \& dangling nodes} 

BikeDNA implements two methods for checking the consistency of network topology at intersections and dead-ends: analysis of undershoots and overshoots (see Fig.~\ref{fig:knownissues}), and analysis of dangling nodes. Under/overshoots commonly occur due to errors in data digitization, but can also be an accurate representation of network conditions, for example when protected bicycle lanes end in intersections that do not provide protection for cyclists. The presence of over- and undershoots skews the ratio of nodes and edges in a network, and thereby distorts network metric computation. Moreover, undershoots in bicycle infrastructure hinder correct routing for cyclists on the network. BikeDNA finds all over/undershoots within a user-defined distance threshold and displays them for further analysis. 

Dangling nodes are nodes of degree one, i.e.~nodes that have only one single edge attached to them. Most infrastructure networks will naturally contain a number of dangling nodes at actual dead-ends or at the endpoints of certain features, e.g.~when a bicycle path ends in the middle of a road. However, they can also be a consequence of under/overshoots or data omissions. It is important to understand whether dangling nodes are caused by actual dead-ends or by digitization errors. BikeDNA therefore visualizes individual dangling nodes and their density on grid cell level to allow investigation of their spatial heterogeneity.

\begin{figure}[t]
\centering
\includegraphics[width=0.999\textwidth]{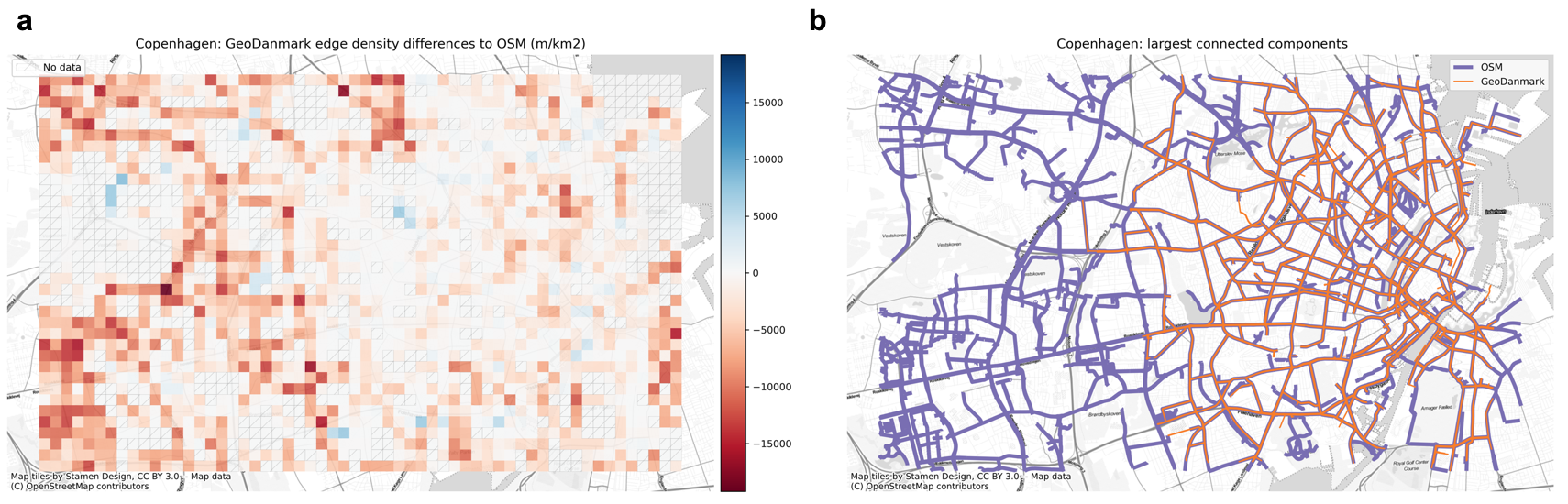}
\caption{Extrinsic analysis of differences between OSM and reference data: a) Comparison of edge density per grid cell. b) Largest connected components. Maps created with BikeDNA v.1.0.0.}
\label{fig:extrinsic_diff}
\end{figure}

\subsection{Network topology: Disconnected components} 

A network component is a maximal set of nodes which are linked by paths. In other words, all nodes within a component can reach each other, but they cannot reach any nodes in the rest of the network. Most real-world bicycle infrastructure networks consist of many disconnected components \citep{natera_orozco_data-driven_2020}. Two disconnected components that are very close to each other can however be a sign of a real `missing link' \citep{vybornova_automated_2021} or of a digitizing error similar to an undershoot. BikeDNA identifies and visualizes all disconnected components of the input network and plots the distribution of all network component lengths on a Zipf plot, which ranks the lengths of all components by descending order. When a Zipf plot follows a straight line in log-log scale, it means that there is a much higher chance to find small disconnected components than expected from traditional exponential distributions \citep{clauset_power-law_2009}. This can mean that there has been no consolidation of the network and elements have been added only piece-wise or randomly \citep{szell_growing_2022}. However, it can also happen that the largest connected component (see the leftmost marker in Fig.~\ref{fig:extrinsic}a at rank $10^0=1$) is a clear outlier, while the rest of the plot follows a different shape. This can mean that network consolidation has taken place. In case of a comparison over the same region, as shown in Fig.~\ref{fig:extrinsic}, if one data set shows a clear outlier in its largest connected component while the other data set does not, it can be an indication that the first data set is more complete. In this particular case, the OSM data set (Fig.~\ref{fig:extrinsic}a) is likely more complete than the reference data (Fig.~\ref{fig:extrinsic}b).

Two disconnected components of bicycle infrastructure might of course be connected by other types of infrastructure, if the entire road network is considered. However, erroneous lack of connections between components of dedicated bicycle infrastructure poses a problem for bicycle routing and will lead to misleading or undesirable results when evaluating bicycle accessibility and Level of Traffic Stress \citep{murphy_implementing_2019, wasserman_evaluating_2019}. BikeDNA therefore also identifies potential missing connections between components by finding and visually highlighting edges from disconnected components that are within a user-defined distance threshold from each other. BikeDNA furthermore conducts a component connectivity analysis on local (grid cell) level, visualizing differences in number of cells reachable from each cell. This measure is particularly useful for comparing and quantifying network reach and connectivity between two data sets (see Fig.~\ref{fig:extrinsic}c,d).

\subsection{Feature matching} 

\begin{figure}[t]
\centering
\includegraphics[width=0.999\textwidth]{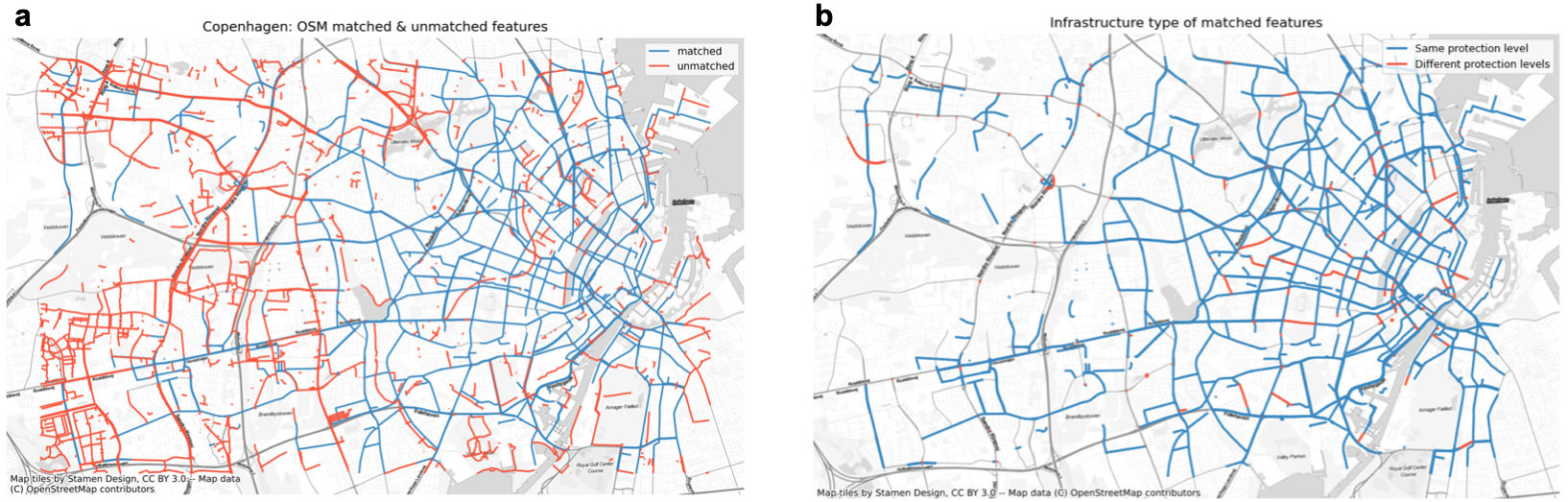}
\caption{Feature matching: a) Matched (blue) and unmatched (red). OSM features b) Matched features with same protection level in both data sets (blue) and differing protection levels (red). Maps created with BikeDNA v.1.0.0.}
\label{fig:featurematching}
\end{figure}

Feature matching is the process of identifying which features from two different data sets correspond to the same real-life object. For example, the same road might be represented by slightly different geometries in two different data sets. Feature matching is necessary for a comparison of \emph{individual features}, rather than feature densities, as well as for transferring attributes from one data set to another, which can be used for example to compensate for missing attribute values. BikeDNA includes a feature matching algorithm which identifies corresponding features between the reference and OSM data sets (see Fig.~\ref{fig:featurematching}). The implemented method converts all network edges to smaller segments of a uniform length before looking for a potential match. The matching algorithm with user-defined input parameters \citep{koukoletsos_assessing_2012, will_development_2014} first identifies all line segments within a maximum search distance of the given edge, and then defines the best match as the line segment(s) that have the smallest undirected Hausdorff distance and the smallest angle in reference to the given edge. 

\subsection{Comparing OSM \& reference data} 

If a reference data set is provided, BikeDNA compares and contrasts it with OSM data from the same area, highlighting how and where the two data sets differ, i.e.~both \textit{how much} bicycle infrastructure is mapped in the two data sets and  \textit{how} the infrastructure is mapped. The comparative analysis makes no prior assumptions about which data set is of higher quality. BikeDNA does thus not lead to an automatic conclusion, but instead requires interpretation of the differences, e.g.~whether differing features are results of errors of omission or commission, and which data set is more fit for purpose. The differences between the data sets are computed and presented on a global and local level, and overlaid where visually adequate (see Fig.~\ref{fig:extrinsic} and Fig.~\ref{fig:extrinsic_diff}). The identified differences can be used to assess the quality of the OSM and reference data sets, and support the decision of which data set to use.

\section{Discussion}

In the following section we sum up BikeDNA’s findings for our showcase area in Greater Copenhagen, and present the general contributions of BikeDNA. We end the section with a discussion of BikeDNA’s limitations and recommendations for future work.

\subsection{Substantial problems with data quality persist}

The Greater Copenhagen area use case demonstrates that data quality issues can result in misleading representation of actual bicycle conditions, both in terms of the extent and the connectivity of the bicycle network. Despite of Denmark being known for its relatively strong bicycle culture \citep{agervig_cycling_2012}, our analysis found several serious issues with bicycle infrastructure data quality, particularly in the administrative reference data set, with large areas being significantly under-mapped. Both OSM and the reference data were moreover suffering from missing links and undershoots resulting in data sets which are substantially more fragmented than the actual bicycle network. Importantly, the analysis indicated large spatial variations in missing or misleading data, which highlights the importance of localized data assessments.

\subsection{BikeDNA: A tool for bicycle researchers, planners, and cyclists}

BikeDNA provides fundamentally new insights into bicycle infrastructure data quality by enabling straightforward exploration of spatial heterogeneity in data quality, and by implementing network-based measures that go beyond quality indicators traditionally applied on bicycle infrastructure data. These novel features open up a plethora of use cases, such as:

\begin{itemize}
\item Urban and regional planning of bicycle infrastructure, where considering the network as a whole is particularly important \citep{szell_growing_2022}.
\item Maintenance and improvement of OSM and administrative data.
\item Transport research on active mobility as well as on multi-modal networks which require high data quality on all transport layers \citep{alessandretti_multimodal_2022}.
\item Improvements to bicycle network data used in tools for transport planning and research, such as `Propensity to Cycle' for transport planning \citep{lovelace_propensity_2017}, `A/B Street' for traffic simulation \citep{carlino_b_2023}, or People for Bike's Bicycle Network Analysis \citep{peopleforbikes_bna_2023}.
\item Citizen science and activism, by enabling citizens' contributions to reliable and quality-assessed data sets, along the lines of other projects supporting citizen data collection for more sustainable transport such as \cite{openbikesensor_openbikesensor_2023} and \cite{streetcomplete_streetcomplete_2023}.
\end{itemize}

\subsection{Limitations}

Although we attempted to cover the main aspects of data quality relevant to bicycle networks, there are some limitations in the design of BikeDNA. 

In terms of data modelling, for the sake of simplicity, BikeDNA makes use of an undirected, simplified network. This means that information about allowed travel directions and turn restrictions is not considered, movement is assumed in both directions on all edges, and different travel directions on the same road are not represented by separate edges. Therefore, the current state of the tool does not make use of routing on the network. For future iterations, however, it might be useful to include travel directions and the underlying road network for accurate path computations.

Another limitation touches upon the core purpose of the tool and the type of results it can produce: since we do not operate with one data set as ground truth against which another data set is evaluated, we cannot conclude where the error lies when differences are identified. For a successful application of BikeDNA, the user is thus expected to have some familiarity with OSM data structure and tagging conventions, but also to have enough knowledge of the study area to correctly interpret the results.

Lastly, we do not directly evaluate the positional accuracy of neither the OSM or the reference data -- although a certain level of comparative positional accuracy can be deduced from the feature matching. 

\subsection{Future work}

A lot of potential future work remains -- not only for quality assessments of bicycle infrastructure data generally, but also for BikeDNA. 

First, it is yet to be determined whether any of the the analyzed metrics can serve as a more general predictor of data quality, based on correlations with other quality metrics. This would allow for a much faster and simpler assessment and remove the need for reference data comparison and more complex analytical tools. Second, the evaluation of bicycle infrastructure data quality in OSM and other data sources should be considered in connection with broader efforts to standardize the mapping of bicycle infrastructure. One challenge of comparing different data sets is translating between different typologies of bicycle infrastructure. Within BikeDNA, we have solved this ambiguity by only distinguishing between protected and unprotected bicycle infrastructure, but this approach could be refined with better classification schemes. To support better planning and research, more work is needed on how to best represent bicycle conditions in data. 

For BikeDNA, a functionality for allowing an easy comparison and feature matching of two non-OSM data sets remains to be implemented. In addition, for the intrinsic analysis of OSM data, a future inclusion of historical data on contributors and edits can offer a complementary perspective, since these metadata are potential indicators of data quality \citep{neis_street_2012, grochenig_estimating_2014}. We hope to hear from the community about future case studies and will be grateful to receive suggestions for possible modifications and improvements to the current tool.

\section{Conclusion}

BikeDNA is an open-source tool for reproducible quality assessments of bicycle infrastructure data. BikeDNA allows for a customized evaluation based on the idea of `fitness of purpose' and comes with a configurable design that can be adapted to different scenarios of data availability, using both stand-alone and comparative methods. The tool computes a wide range of quality metrics and extends previous research on bicycle infrastructure quality by adding a focus on data topology and structural properties. The provided example application of BikeDNA based on a comparison of OSM and Danish administrative data has demonstrated that BikeDNA can reveal important spatial variations, e.g.~in missing data and topology errors; and that quality assessments of bicycle infrastructure data are necessary, as demonstrated by the many inconsistencies and errors revealed by the example analysis.
Some open questions remain, both for BikeDNA and more broadly, regarding the definition and collection of high-quality data on bicycle infrastructure. The absence of high-quality official data and recognized typologies for bicycle infrastructure reflects how active mobility modes have been historically neglected. For cities, regions and countries around the world that strive to make use of recent advances in data-driven planning to expand their bicycle networks, better data on existing bicycle conditions are urgently needed.

\begin{acks}
Thanks to all OSM contributors, whose efforts make spatial data open and free, to those developing and contributing to the Python libraries making this work possible, and to Clément Sebastiao for developing the modified OSMnx function used for network simplification. We acknowledge support by the Danish Ministry of Transport.
\end{acks}


\bibliographystyle{SageH}
\bibliography{paper.bib}

\begin{thebibliography}{54}
\providecommand{\natexlab}[1]{#1}
\providecommand{\url}[1]{\texttt{#1}}
\providecommand{\urlprefix}{URL }
\expandafter\ifx\csname urlstyle\endcsname\relax
  \providecommand{\doi}[1]{DOI:\discretionary{}{}{}#1}\else
  \providecommand{\doi}{DOI:\discretionary{}{}{}\begingroup
  \urlstyle{rm}\Url}\fi

\bibitem[{Agervig and Ebert(2012)}]{agervig_cycling_2012}
Agervig CT and Ebert AK (2012) Cycling {Cultures} in {Northern} {Europe}:
  {From} ‘{Golden} {Age}’ to ‘{Renaissance}’.
\newblock In: Parkin J (ed.) \emph{Cycling and {Sustainability}},
  \emph{Transport and {Sustainability}}, volume~1. Emerald Group Publishing
  Limited.
\newblock ISBN 978-1-78052-299-9 978-1-78052-298-2, pp. 23--58.
\newblock \doi{10.1108/S2044-9941(2012)0000001004}.
\newblock \urlprefix\url{https://doi.org/10.1108/S2044-9941(2012)0000001004}.

\bibitem[{Alessandretti et~al.(2022)Alessandretti, Natera~Orozco, Saberi, Szell
  and Battiston}]{alessandretti_multimodal_2022}
Alessandretti L, Natera~Orozco LG, Saberi M, Szell M and Battiston F (2022)
  Multimodal urban mobility and multilayer transport networks.
\newblock \emph{Environment and Planning B: Urban Analytics and City Science} :
  239980832211081\doi{10.1177/23998083221108190}.
\newblock
  \urlprefix\url{http://journals.sagepub.com/doi/10.1177/23998083221108190}.

\bibitem[{Ballatore and Zipf(2015)}]{fabrikant_conceptual_2015}
Ballatore A and Zipf A (2015) A {Conceptual} {Quality} {Framework} for
  {Volunteered} {Geographic} {Information}.
\newblock In: Fabrikant SI, Raubal M, Bertolotto M, Davies C, Freundschuh S and
  Bell S (eds.) \emph{Spatial {Information} {Theory}}, volume 9368. Cham:
  Springer International Publishing.
\newblock ISBN 978-3-319-23373-4 978-3-319-23374-1, pp. 89--107.
\newblock \doi{10.1007/978-3-319-23374-1_5}.
\newblock \urlprefix\url{http://link.springer.com/10.1007/978-3-319-23374-1_5}.
\newblock Series Title: Lecture Notes in Computer Science.

\bibitem[{Barron et~al.(2014)Barron, Neis and Zipf}]{barron_comprehensive_2014}
Barron C, Neis P and Zipf A (2014) A {Comprehensive} {Framework} for
  {Intrinsic} {OpenStreetMap} {Quality} {Analysis}: {A} {Comprehensive}
  {Framework} for {Intrinsic} {OpenStreetMap} {Quality} {Analysis}.
\newblock \emph{Transactions in GIS} 18(6): 877--895.
\newblock \doi{10.1111/tgis.12073}.
\newblock
  \urlprefix\url{https://onlinelibrary.wiley.com/doi/10.1111/tgis.12073}.

\bibitem[{Boeing(2017)}]{boeing_osmnx_2017}
Boeing G (2017) {OSMnx}: {New} methods for acquiring, constructing, analyzing,
  and visualizing complex street networks.
\newblock \emph{Computers, Environment and Urban Systems} 65: 126--139.
\newblock \doi{10.1016/j.compenvurbsys.2017.05.004}.
\newblock
  \urlprefix\url{https://linkinghub.elsevier.com/retrieve/pii/S0198971516303970}.

\bibitem[{Brovelli et~al.(2017)Brovelli, Minghini, Molinari and
  Mooney}]{brovelli_towards_2017}
Brovelli MA, Minghini M, Molinari M and Mooney P (2017) Towards an {Automated}
  {Comparison} of {OpenStreetMap} with {Authoritative} {Road} {Datasets}.
\newblock \emph{Transactions in GIS} 21(2): 191--206.
\newblock \doi{10.1111/tgis.12182}.
\newblock
  \urlprefix\url{https://onlinelibrary.wiley.com/doi/10.1111/tgis.12182}.

\bibitem[{{C40 Cities}(2019)}]{c40_cities_upgrade_2019}
{C40 Cities} (2019) Upgrade of the {Cycle} {Network} in {Bogotà}
  {Dramatically} {Increases} {Bike} {Trips}.
\newblock
  \urlprefix\url{https://www.c40.org/case-studies/upgrade-of-the-cycle-network-in-bogota-dramatically-increases-bike-trips/}.

\bibitem[{Carlino et~al.(2023)Carlino, Li and Kirk}]{carlino_b_2023}
Carlino D, Li Y and Kirk M (2023) A/{B} {Street}.
\newblock \urlprefix\url{https://github.com/a-b-street/abstreet}.
\newblock Original-date: 2018-06-04T00:44:43Z.

\bibitem[{{City Of Paris}(2021)}]{city_of_paris_nouveau_2021}
{City Of Paris} (2021) Un nouveau plan vélo pour une ville 100 \% cyclable.
\newblock
  \urlprefix\url{https://www.paris.fr/pages/un-nouveau-plan-velo-pour-une-ville-100-cyclable-19554}.

\bibitem[{Clauset et~al.(2009)Clauset, Shalizi and
  Newman}]{clauset_power-law_2009}
Clauset A, Shalizi CR and Newman MEJ (2009) Power-{Law} {Distributions} in
  {Empirical} {Data}.
\newblock \emph{SIAM Review} 51(4): 661--703.
\newblock \doi{10.1137/070710111}.
\newblock \urlprefix\url{https://doi.org/10.1137/070710111}.

\bibitem[{Degrossi et~al.(2018)Degrossi, Porto~de Albuquerque, Santos~Rocha and
  Zipf}]{degrossi_taxonomy_2018}
Degrossi LC, Porto~de Albuquerque J, Santos~Rocha Rd and Zipf A (2018) A
  taxonomy of quality assessment methods for volunteered and crowdsourced
  geographic information.
\newblock \emph{Transactions in GIS} 22(2): 542--560.
\newblock \doi{10.1111/tgis.12329}.
\newblock
  \urlprefix\url{https://onlinelibrary.wiley.com/doi/abs/10.1111/tgis.12329}.

\bibitem[{Dill(2004)}]{dill_measuring_2004}
Dill J (2004) Measuring {Network} {Connectivity} for {Bicycling} and {Walking}.
\newblock In: \emph{83rd {Annual} {Meeting} of the {Transportation} {Research}
  {Board}}, volume~83. Washington, DC, p.~19.

\bibitem[{EC(2021)}]{ec_new_2021}
EC (2021) The {New} {EU} {Urban} {Mobility} {Framework}.
\newblock Technical report, European Commission, Brussels.
\newblock
  \urlprefix\url{https://transport.ec.europa.eu/system/files/2021-12/com_2021_811_the-new-eu-urban-mobility.pdf}.

\bibitem[{Ferster et~al.(2020)Ferster, Fischer, Manaugh, Nelson and
  Winters}]{ferster_using_2020}
Ferster C, Fischer J, Manaugh K, Nelson T and Winters M (2020) Using
  {OpenStreetMap} to inventory bicycle infrastructure: {A} comparison with open
  data from cities.
\newblock \emph{International Journal of Sustainable Transportation} 14(1):
  64--73.
\newblock \doi{10.1080/15568318.2018.1519746}.
\newblock
  \urlprefix\url{https://www.tandfonline.com/doi/full/10.1080/15568318.2018.1519746}.

\bibitem[{Fleischmann(2019)}]{fleischmann_momepy_2019}
Fleischmann M (2019) momepy: {Urban} {Morphology} {Measuring} {Toolkit}.
\newblock \emph{Journal of Open Source Software} 4(43): 1807.
\newblock \doi{10.21105/joss.01807}.
\newblock \urlprefix\url{https://joss.theoj.org/papers/10.21105/joss.01807}.

\bibitem[{Fonte(2017)}]{fonte_assessing_2017}
Fonte C (2017) Assessing {VGI} {Data} {Quality}.
\newblock \emph{Mapping and the Citizen Sensor} : 137--163\doi{10.5334/bbf.g}.
\newblock
  \urlprefix\url{https://www.ubiquitypress.com/site/chapters/10.5334/bbf.g/}.

\bibitem[{GeoDanmark(2023)}]{geodanmark_2023}
GeoDanmark (2023) Danmarks {Geografi} - {GeoDanmark}.
\newblock \urlprefix\url{https://dataforsyningen.dk/data/3563}.

\bibitem[{Girres and Touya(2010)}]{girres_quality_2010}
Girres JF and Touya G (2010) Quality {Assessment} of the {French}
  {OpenStreetMap} {Dataset}: {Quality} {Assessment} of the {French}
  {OpenStreetMap} {Dataset}.
\newblock \emph{Transactions in GIS} 14(4): 435--459.
\newblock \doi{10.1111/j.1467-9671.2010.01203.x}.
\newblock
  \urlprefix\url{https://onlinelibrary.wiley.com/doi/10.1111/j.1467-9671.2010.01203.x}.

\bibitem[{Graser et~al.(2014)Graser, Straub and
  Dragaschnig}]{graser_towards_2014}
Graser A, Straub M and Dragaschnig M (2014) Towards an {Open} {Source}
  {Analysis} {Toolbox} for {Street} {Network} {Comparison}: {Indicators},
  {Tools} and {Results} of a {Comparison} of {OSM} and the {Official}
  {Austrian} {Reference} {Graph}.
\newblock \emph{Transactions in GIS} 18(4): 510--526.
\newblock \doi{10.1111/tgis.12061}.
\newblock
  \urlprefix\url{https://search.ebscohost.com/login.aspx?direct=true&db=bth&AN=97351436&site=ehost-live}.
\newblock Publisher: Wiley-Blackwell.

\bibitem[{Gröchenig et~al.(2014)Gröchenig, Brunauer and
  Rehrl}]{grochenig_estimating_2014}
Gröchenig S, Brunauer R and Rehrl K (2014) Estimating {Completeness} of {VGI}
  {Datasets} by {Analyzing} {Community} {Activity} {Over} {Time} {Periods}.
\newblock In: Huerta J, Schade S and Granell C (eds.) \emph{Connecting a
  {Digital} {Europe} {Through} {Location} and {Place}}, Lecture {Notes} in
  {Geoinformation} and {Cartography}. Cham: Springer International Publishing.
\newblock ISBN 978-3-319-03611-3, pp. 3--18.
\newblock \doi{10.1007/978-3-319-03611-3_1}.
\newblock \urlprefix\url{https://doi.org/10.1007/978-3-319-03611-3_1}.

\bibitem[{Hagberg et~al.(2008)Hagberg, Schult and
  Swart}]{hagberg_exploring_2008}
Hagberg AA, Schult DA and Swart PJ (2008) Exploring {Network} {Structure},
  {Dynamics}, and {Function} using {NetworkX}.
\newblock In: Varoquaux G, Vaught T and Millman J (eds.) \emph{Proceedings of
  the 7th {Python} in {Science} {Conference}}. Pasadena, CA USA, pp. 11 -- 15.

\bibitem[{Haklay(2010)}]{haklay_how_2010-1}
Haklay M (2010) How {Good} is {Volunteered} {Geographical} {Information}? {A}
  {Comparative} {Study} of {OpenStreetMap} and {Ordnance} {Survey} {Datasets}.
\newblock \emph{Environment and Planning B: Planning and Design} 37(4):
  682--703.
\newblock \doi{10.1068/b35097}.
\newblock \urlprefix\url{http://journals.sagepub.com/doi/10.1068/b35097}.

\bibitem[{Haklay et~al.(2010)Haklay, Basiouka, Antoniou and
  Ather}]{haklay_how_2010}
Haklay MM, Basiouka S, Antoniou V and Ather A (2010) How {Many} {Volunteers}
  {Does} it {Take} to {Map} an {Area} {Well}? {The} {Validity} of {Linus}’
  {Law} to {Volunteered} {Geographic} {Information}.
\newblock \emph{The Cartographic Journal} 47(4): 315--322.
\newblock \doi{10.1179/000870410X12911304958827}.
\newblock \urlprefix\url{https://doi.org/10.1179/000870410X12911304958827}.

\bibitem[{Hashemi and Abbaspour(2015)}]{hashemi_assessment_2015}
Hashemi P and Abbaspour RA (2015) Assessment of {Logical} {Consistency} in
  {OpenStreetMap} {Based} on the {Spatial} {Similarity} {Concept}.
\newblock In: Jokar~Arsanjani J, Zipf A, Mooney P and Helbich M (eds.)
  \emph{{OpenStreetMap} in {GIScience}: {Experiences}, {Research}, and
  {Applications}}, Lecture {Notes} in {Geoinformation} and {Cartography}. Cham:
  Springer International Publishing.
\newblock ISBN 978-3-319-14280-7, pp. 19--36.
\newblock \doi{10.1007/978-3-319-14280-7_2}.
\newblock \urlprefix\url{https://doi.org/10.1007/978-3-319-14280-7_2}.

\bibitem[{Hochmair et~al.(2015)Hochmair, Zielstra and
  Neis}]{hochmair_assessing_2015}
Hochmair HH, Zielstra D and Neis P (2015) Assessing the {Completeness} of
  {Bicycle} {Trail} and {Lane} {Features} in {OpenStreetMap} for the {United}
  {States}: {Completeness} of {Bicycle} {Features} in {OpenStreetMap}.
\newblock \emph{Transactions in GIS} 19(1): 63--81.
\newblock \doi{10.1111/tgis.12081}.
\newblock
  \urlprefix\url{https://onlinelibrary.wiley.com/doi/10.1111/tgis.12081}.

\bibitem[{Hvingel and Jensen(2023)}]{hvingel_gode_2023}
Hvingel L and Jensen T (2023) Gode cykeldata til alle.
\newblock \emph{Trafik og veje}
  \urlprefix\url{https://www.kl.dk/media/53604/artikel_gode_cykeldata_trafik_og_veje_jan_2023_layout.pdf}.

\bibitem[{Jaramillo et~al.(2022)Jaramillo, Kahn~Ribeiro, Newman, Dhar,
  Diemuodeke, Kajino, Lee, Nugroho, Ou, Hammer~Strømman and
  Whitehead}]{jaramillo_transport_2022}
Jaramillo P, Kahn~Ribeiro S, Newman P, Dhar S, Diemuodeke OE, Kajino T, Lee DS,
  Nugroho SB, Ou X, Hammer~Strømman A and Whitehead J (2022) Transport.
\newblock In: \emph{Climate {Change} 2022: {Mitigation} of {Climate} {Change}}.
  Intergovernmental Panel on Climate Change (IPCC), pp. 1049--1160.
\newblock \urlprefix\url{doi:10.1017/9781009157926.012}.

\bibitem[{Jordahl et~al.(2021)Jordahl, Bossche, Fleischmann, McBride,
  Wasserman, Badaracco, Gerard, Snow, Tratner, Perry, Farmer, Hjelle, Cochran,
  Gillies, Culbertson, Bartos, Ward, Caria, Taves, Eubank, sangarshanan,
  Flavin, Richards, Rey, maxalbert, Bilogur, Ren, Arribas-Bel, Mesejo-León and
  Wasser}]{jordahl_geopandasgeopandas_2021}
Jordahl K, Bossche JVd, Fleischmann M, McBride J, Wasserman J, Badaracco AG,
  Gerard J, Snow AD, Tratner J, Perry M, Farmer C, Hjelle GA, Cochran M,
  Gillies S, Culbertson L, Bartos M, Ward B, Caria G, Taves M, Eubank N,
  sangarshanan, Flavin J, Richards M, Rey S, maxalbert, Bilogur A, Ren C,
  Arribas-Bel D, Mesejo-León D and Wasser L (2021) geopandas/geopandas:
  v0.10.2.
\newblock \doi{10.5281/zenodo.5573592}.
\newblock \urlprefix\url{https://zenodo.org/record/5573592}.

\bibitem[{Keßler et~al.(2011)Keßler, Trame and
  Kauppinen}]{kesler_tracking_2011}
Keßler C, Trame J and Kauppinen T (2011) Tracking {Editing} {Processes} in
  {Volunteered} {Geographic} {Information}: {The} {Case} of {OpenStreetMap}.
\newblock \emph{Proceedings of the Conference on Spatial Information Theory,
  Workshop: Identifying Objects, Processes and Events in Spatio-Temporally
  Distributed Data} : 7.

\bibitem[{Koukoletsos et~al.(2012)Koukoletsos, Haklay and
  Ellul}]{koukoletsos_assessing_2012}
Koukoletsos T, Haklay M and Ellul C (2012) Assessing {Data} {Completeness} of
  {VGI} through an {Automated} {Matching} {Procedure} for {Linear} {Data}.
\newblock \emph{Transactions in GIS} 16(4): 477--498.
\newblock \doi{10.1111/j.1467-9671.2012.01304.x}.
\newblock
  \urlprefix\url{https://onlinelibrary.wiley.com/doi/abs/10.1111/j.1467-9671.2012.01304.x}.

\bibitem[{Lovelace et~al.(2017)Lovelace, Goodman, Aldred, Berkoff, Abbas and
  Woodcock}]{lovelace_propensity_2017}
Lovelace R, Goodman A, Aldred R, Berkoff N, Abbas A and Woodcock J (2017) The
  {Propensity} to {Cycle} {Tool}: {An} open source online system for
  sustainable transport planning.
\newblock \emph{Journal of Transport and Land Use} 10(1).
\newblock \doi{10.5198/jtlu.2016.862}.
\newblock \urlprefix\url{https://www.jtlu.org/index.php/jtlu/article/view/862}.
\newblock Number: 1.

\bibitem[{Medeiros and Holanda(2019)}]{medeiros_solutions_2019}
Medeiros G and Holanda M (2019) Solutions for {Data} {Quality} in {GIS} and
  {VGI}: {A} {Systematic} {Literature} {Review}.
\newblock In: Rocha A, Adeli H, Reis LP and Costanzo S (eds.) \emph{New
  {Knowledge} in {Information} {Systems} and {Technologies}}, Advances in
  {Intelligent} {Systems} and {Computing}. Cham: Springer International
  Publishing.
\newblock ISBN 978-3-030-16181-1, pp. 645--654.
\newblock \doi{10.1007/978-3-030-16181-1_61}.

\bibitem[{Mekuria et~al.(2012)Mekuria, Furth and
  Nixon}]{mekuria_low-stress_2012}
Mekuria MC, Furth PG and Nixon H (2012) Low-{Stress} {Bicycling} and {Network}
  {Connectivity}.
\newblock Technical Report 11-19, Mineta Transportation Institute.
\newblock
  \urlprefix\url{https://www.semanticscholar.org/paper/Low-Stress-Bicycling-and-Network-Connectivity-Mekuria-Furth/a50063c06112d3eb6aa752dfd362e1bdbc7f1c7e}.

\bibitem[{Mondzech and Sester(2011)}]{mondzech_quality_2011}
Mondzech J and Sester M (2011) Quality {Analysis} of {OpenStreetMap} {Data}
  {Based} on {Application} {Needs}.
\newblock \emph{Cartographica: The International Journal for Geographic
  Information and Geovisualization} 46(2): 115--125.
\newblock \doi{10.3138/carto.46.2.115}.
\newblock
  \urlprefix\url{https://www.utpjournals.press/doi/abs/10.3138/carto.46.2.115}.
\newblock Publisher: University of Toronto Press.

\bibitem[{Murphy and Owen(2019)}]{murphy_implementing_2019}
Murphy B and Owen A (2019) Implementing {Low}-{Stress} {Bicycle} {Routing} in
  {National} {Accessibility} {Evaluation}.
\newblock \emph{Transportation Research Record} 2673(5): 240--249.
\newblock \doi{10.1177/0361198119837179}.
\newblock \urlprefix\url{https://doi.org/10.1177/0361198119837179}.
\newblock Publisher: SAGE Publications Inc.

\bibitem[{Natera~Orozco et~al.(2020)Natera~Orozco, Battiston, Iñiguez and
  Szell}]{natera_orozco_data-driven_2020}
Natera~Orozco LG, Battiston F, Iñiguez G and Szell M (2020) Data-driven
  strategies for optimal bicycle network growth.
\newblock \emph{Royal Society Open Science} 7(12): 201130.
\newblock \doi{10.1098/rsos.201130}.
\newblock
  \urlprefix\url{https://royalsocietypublishing.org/doi/full/10.1098/rsos.201130}.
\newblock Publisher: Royal Society.

\bibitem[{Neis et~al.(2012)Neis, Zielstra and Zipf}]{neis_street_2012}
Neis P, Zielstra D and Zipf A (2012) The {Street} {Network} {Evolution} of
  {Crowdsourced} {Maps}: {OpenStreetMap} in {Germany} 2007–2011.
\newblock \emph{Future Internet} 4(1): 1--21.
\newblock \doi{10.3390/fi4010001}.
\newblock \urlprefix\url{https://www.mdpi.com/1999-5903/4/1/1}.
\newblock Number: 1 Publisher: Molecular Diversity Preservation International.

\bibitem[{Neis et~al.(2013)Neis, Zielstra and Zipf}]{neis_comparison_2013}
Neis P, Zielstra D and Zipf A (2013) Comparison of {Volunteered} {Geographic}
  {Information} {Data} {Contributions} and {Community} {Development} for
  {Selected} {World} {Regions}.
\newblock \emph{Future Internet} 5(2): 282--300.
\newblock \doi{10.3390/fi5020282}.
\newblock \urlprefix\url{https://www.mdpi.com/1999-5903/5/2/282}.
\newblock Number: 2 Publisher: Multidisciplinary Digital Publishing Institute.

\bibitem[{Nelson et~al.(2021)Nelson, Ferster, Laberee, Fuller and
  Winters}]{nelson_crowdsourced_2021}
Nelson T, Ferster C, Laberee K, Fuller D and Winters M (2021) Crowdsourced data
  for bicycling research and practice.
\newblock \emph{Transport Reviews} 41(1): 97--114.
\newblock \doi{10.1080/01441647.2020.1806943}.
\newblock \urlprefix\url{https://doi.org/10.1080/01441647.2020.1806943}.

\bibitem[{Olmos et~al.(2020)Olmos, Tadeo, Vlachogiannis, Alhasoun,
  Espinet~Alegre, Ochoa, Targa and González}]{olmos_data_2020}
Olmos LE, Tadeo MS, Vlachogiannis D, Alhasoun F, Espinet~Alegre X, Ochoa C,
  Targa F and González MC (2020) A data science framework for planning the
  growth of bicycle infrastructures.
\newblock \emph{Transportation Research Part C: Emerging Technologies} 115:
  102640.
\newblock \doi{10.1016/j.trc.2020.102640}.
\newblock
  \urlprefix\url{https://www.sciencedirect.com/science/article/pii/S0968090X19306436}.

\bibitem[{OpenBikeSensor(2023)}]{openbikesensor_openbikesensor_2023}
OpenBikeSensor (2023) {OpenBikeSensor}.
\newblock \urlprefix\url{https://openbikesensor.org/en/}.

\bibitem[{PeopleForBikes(2023)}]{peopleforbikes_bna_2023}
PeopleForBikes (2023) {BNA} {Bicycle} {Network} {Analysis}.
\newblock \urlprefix\url{https://bna.peopleforbikes.org/#/}.

\bibitem[{{Rambøll}(2022)}]{ramboll_walking_2022}
{Rambøll} (2022) Walking and cycling data. {Practice}, challenges, needs and
  gaps.
\newblock
  \urlprefix\url{https://ramboll.com/-/media/files/rgr/documents/markets/transport/walking-cycling-data-gaps-2022.pdf}.

\bibitem[{Sebastiao(2022)}]{sebastiao_modified_2022}
Sebastiao C (2022) Modified version of {OSMnx} {Simplification}.
\newblock \emph{GitHub} \urlprefix\url{https://github.com/anerv/NERDS_osmnx}.

\bibitem[{Senaratne et~al.(2017)Senaratne, Mobasheri, Ali, Capineri and
  Haklay}]{senaratne_review_2017}
Senaratne H, Mobasheri A, Ali AL, Capineri C and Haklay MM (2017) A review of
  volunteered geographic information quality assessment methods.
\newblock \emph{International Journal of Geographical Information Science}
  31(1): 139--167.
\newblock \doi{10.1080/13658816.2016.1189556}.
\newblock
  \urlprefix\url{https://www.tandfonline.com/doi/full/10.1080/13658816.2016.1189556}.

\bibitem[{Steinacker et~al.(2022)Steinacker, Storch, Timme and
  Schröder}]{steinacker_demand-driven_2022}
Steinacker C, Storch DM, Timme M and Schröder M (2022) Demand-driven design of
  bicycle infrastructure networks for improved urban bikeability.
\newblock \emph{Nature Computational Science} :
  1--10\doi{10.1038/s43588-022-00318-w}.
\newblock \urlprefix\url{https://www.nature.com/articles/s43588-022-00318-w}.
\newblock Publisher: Nature Publishing Group.

\bibitem[{StreetComplete(2023)}]{streetcomplete_streetcomplete_2023}
StreetComplete (2023) {StreetComplete}.
\newblock \urlprefix\url{https://streetcomplete.app}.

\bibitem[{Szell et~al.(2022)Szell, Mimar, Perlman, Ghoshal and
  Sinatra}]{szell_growing_2022}
Szell M, Mimar S, Perlman T, Ghoshal G and Sinatra R (2022) Growing urban
  bicycle networks.
\newblock \emph{Scientific Reports} 12(1): 6765.
\newblock \doi{10.1038/s41598-022-10783-y}.
\newblock \urlprefix\url{https://www.nature.com/articles/s41598-022-10783-y}.
\newblock Number: 1 Publisher: Nature Publishing Group.

\bibitem[{Vybornova et~al.(2022)Vybornova, Cunha, Gühnemann and
  Szell}]{vybornova_automated_2021}
Vybornova A, Cunha T, Gühnemann A and Szell M (2022) Automated {Detection} of
  {Missing} {Links} in {Bicycle} {Networks}.
\newblock \emph{Geographical Analysis} n/a(n/a).
\newblock \doi{10.1111/gean.12324}.
\newblock
  \urlprefix\url{https://onlinelibrary.wiley.com/doi/abs/10.1111/gean.12324}.

\bibitem[{Wasserman et~al.(2019)Wasserman, Rixey, Zhou, Levitt and
  Benjamin}]{wasserman_evaluating_2019}
Wasserman D, Rixey A, Zhou XE, Levitt D and Benjamin M (2019) Evaluating
  {OpenStreetMap}’s {Performance} {Potential} for {Level} of {Traffic}
  {Stress} {Analysis}.
\newblock \emph{Transportation Research Record} 2673(4): 284--294.
\newblock \doi{10.1177/0361198119836772}.
\newblock \urlprefix\url{https://doi.org/10.1177/0361198119836772}.
\newblock Publisher: SAGE Publications Inc.

\bibitem[{Will(2014)}]{will_development_2014}
Will J (2014) \emph{Development of an automated matching algorithm to assess
  the quality of the {OpenStreetMap} road network: a case study in {Göteborg},
  {Sweden}}.
\newblock PhD Thesis, Lund University.
\newblock
  \urlprefix\url{https://www.semanticscholar.org/paper/Development-of-an-automated-matching-algorithm-to-%3A-Will/b3b77d579077b967820630db56522bef31654f21}.

\bibitem[{Zhang et~al.(2021)Zhang, Wang, Jiao, Zhou, Yu and
  Cheng}]{zhang_detecting_2021}
Zhang X, Wang T, Jiao D, Zhou Z, Yu J and Cheng X (2021) Detecting inconsistent
  information in crowd-sourced street networks based on parallel carriageways
  identification and the rule of symmetry.
\newblock \emph{ISPRS Journal of Photogrammetry and Remote Sensing} 175:
  386--402.
\newblock \doi{10.1016/j.isprsjprs.2021.03.014}.
\newblock
  \urlprefix\url{https://www.sciencedirect.com/science/article/pii/S092427162100085X}.

\bibitem[{Zielstra et~al.(2013)Zielstra, Hochmair and
  Neis}]{zielstra_assessing_2013}
Zielstra D, Hochmair H and Neis P (2013) Assessing the {Effect} of {Data}
  {Imports} on the {Completeness} of {OpenStreetMap} – {A} {United} {States}
  {Case} {Study}.
\newblock \emph{Transactions in GIS} 17.
\newblock \doi{10.1111/tgis.12037}.

\bibitem[{Zielstra and Hochmair(2012)}]{zielstra_using_2012}
Zielstra D and Hochmair HH (2012) Using {Free} and {Proprietary} {Data} to
  {Compare} {Shortest}-{Path} {Lengths} for {Effective} {Pedestrian} {Routing}
  in {Street} {Networks}.
\newblock \emph{Transportation Research Record} 2299(1): 41--47.
\newblock \doi{10.3141/2299-05}.
\newblock \urlprefix\url{https://doi.org/10.3141/2299-05}.
\newblock Publisher: SAGE Publications Inc.

\end{thebibliography}

\end{document}